\documentclass{Interspeech}

\interspeechcameraready

\title{Developing a Top-tier Framework in Naturalistic Conditions Challenge for Categorized Emotion Prediction: From Speech Foundation Models and Learning Objective to Data Augmentation and Engineering Choices}

\author[affiliation={1}]{*Tiantian}{Feng} 
\author[affiliation={1}]{*Thanathai}{Lertpetchpun}
\author[affiliation={1}]{Dani}{Byrd}
\author[affiliation={1}]{Shrikanth}{Narayanan}

\affiliation{University of Southern California}{Los Angeles}{USA}
\email{tiantiaf, lertpetc@usc.edu}

\usepackage{lipsum}

\newcommand\blfootnote[1]{%
  \begingroup
  \renewcommand\thefootnote{}\footnote{#1}%
  \addtocounter{footnote}{-1}%
  \endgroup
}

\usepackage{comment}
\usepackage{multirow}
\usepackage{xcolor}
\usepackage{booktabs}
\usepackage{hyperref}
\begin{document}

\maketitle

\begin{abstract}
Speech emotion recognition (SER), particularly for naturally expressed emotions, remains a challenging computational task. Key challenges include the inherent subjectivity in emotion annotation and the imbalanced distribution of emotion labels in datasets. This paper introduces the \texttt{SAILER} system developed for participation in the INTERSPEECH 2025 Emotion Recognition Challenge (Task 1). The challenge dataset, which contains natural emotional speech from podcasts, serves as a valuable resource for studying imbalanced and subjective emotion annotations. Our system is designed to be simple, reproducible, and effective, highlighting critical choices in modeling, learning objectives, data augmentation, and engineering choices. Results show that even a single system (without ensembling) can outperform more than 95\% of the submissions, with a Macro-F1 score exceeding 0.4. Moreover, an ensemble of three systems further improves performance, achieving a competitively ranked score (top-3 performing team). Our model is at: \href{https://github.com/tiantiaf0627/vox-profile-release}{https://github.com/tiantiaf0627/vox-profile-release}.
\vspace{0.5mm}
\end{abstract}

\section{Introduction}
\blfootnote{*indicates equal contribution}
Despite notable advances in speech emotion recognition (SER) driven by speech foundation models \cite{bommasani2021opportunities, wagner2023dawn}, accurately recognizing emotions from speech remains a challenging computational problem in machine learning \cite{lee2023engineering}. Key challenges include inherent ambiguity in emotion expressions \cite{Mower2009Interpretingambiguousemotionalexpressions} and subjectivity in annotating emotions \cite{chien2024balancing,booth2024people}, imbalanced distributions in emotion labels, and complexity of speaker (and listener) environments and contexts. For example, the previous Odyssey-Speech Emotion Challenge reports that the top-performing team achieved only around 0.35 macro-F1 score in an 8-emotion classification problem \cite{goncalves2024odyssey, chen24_odyssey}. Previous approaches have commonly employed techniques such as focal loss to mitigate class imbalance \cite{costa24_odyssey, chen24_odyssey}. Moreover, a universally deployed approach is used to leverage recently developed speech foundation models, which have shown promising improvements in SER performance \cite{goncalves2024odyssey}.

This paper introduces a promising solution to Task 1 of the IS25-Speech Emotion Recognition in Naturalistic Conditions Challenge (which we refer to as the IS25-SER Challenge) \cite{Naini_2025}. Apart from building on the success of the past Odyssey-SER Challenge, we identified several elements to target for improvement in this IS25-SER challenge~\footnote{\href{https://lab-msp.com/MSP-Podcast_Competition/IS2025/}{https://lab-msp.com/MSP-Podcast\_Competition/IS2025/}} \cite{goncalves2024odyssey}. First, the predominant approach to modeling SER relies on hard labeling (one-hot encoding), often ignoring the complete array of annotations available in the dataset. This limitation leads many approaches to discard samples lacking annotation agreement during training, resulting in significant data loss. Second, data augmentation, a simple technique for diversifying input data, has rarely been explored to mitigate data imbalance in previous challenges. Finally, many top-performing teams opt to ensemble five or more models in their final solutions, making reproducibility practically unfriendly for the research community. Therefore, we focus on designing a reproducibility-friendly solution with minimum complexity and choice of hyperparameters.

We introduce SAIL-Emotion Recognition (\texttt{SAILER}), a SER framework that systematically explores several key design considerations for effective SER modeling. SAILER addresses various aspects of SER, including modeling choice, learning objectives, data augmentation, and engineering design choices to tackle class imbalances. Our design philosophy prioritizes simplicity, reproducibility, and efficiency, avoiding over-engineered solutions, unnecessary system complexity, and bulky ensembles. We aim for the community to replicate our system with minimal time and effort beyond the challenge. Experimental results show that \textbf{a single system (without any ensemble) can already achieve the top-tier ranks in the leaderboard}, a macro-F1 above 0.4 among over 150 submissions. Furthermore, an ensemble of three systems improves the macro-F1 score above 0.41 (outperforming 95\% submissions). 


Our implementation, along with the best-performing single system, will be publicly released and could be considered as baselines for future research. The key design concepts and findings are summarized below:

\begin{itemize}[leftmargin=*]
    \item A simple concatenation of output from speech and text foundation models is sufficient to achieve top-tier scores.
    
    \item Modeling emotion distribution is more effective than using cross-entropy loss by utilizing more data in the training
    
    \item Simple but novel data augmentation techniques, such as audio mixing and annotation dropout, further improve SER performance, especially for predicting minority emotion classes.
    
    \item Engineering choices, including distribution reweighting, integrating additional validation metrics for minority emotion classes, and predicting \textit{other emotion} attribute labels, are benefit SER performance.
    
\end{itemize}

\section{Method}

\subsection{Speech Foundation Model}
The use of speech foundation models, such as Whisper \cite{radford2023robust} and WavLM \cite{chen2022wavlm}, have demonstrated effectiveness in SER. Many recent studies \cite{pepino21_interspeech, feng2024foundation} show that simply leveraging pre-trained speech representations is adequate to achieve competitive performance compared to traditional hand-crafted features. Furthermore, our literature review indicates that downstream architecture has a minimum impact on SER performance. Consequently, the \texttt{SAILER} system adopts the simple downstream model described in \cite{feng2023peft}. Specifically, the model processes the last hidden output or a weighted average of all hidden outputs from the encoder layers. This representation is passed through a 3-layer pointwise convolutional module, followed by temporal averaging. Finally, the averaged output is fed into a two-layer MLP with ReLU activation functions in between.

\begin{figure}
    \centering
    \includegraphics[width=0.68\linewidth]{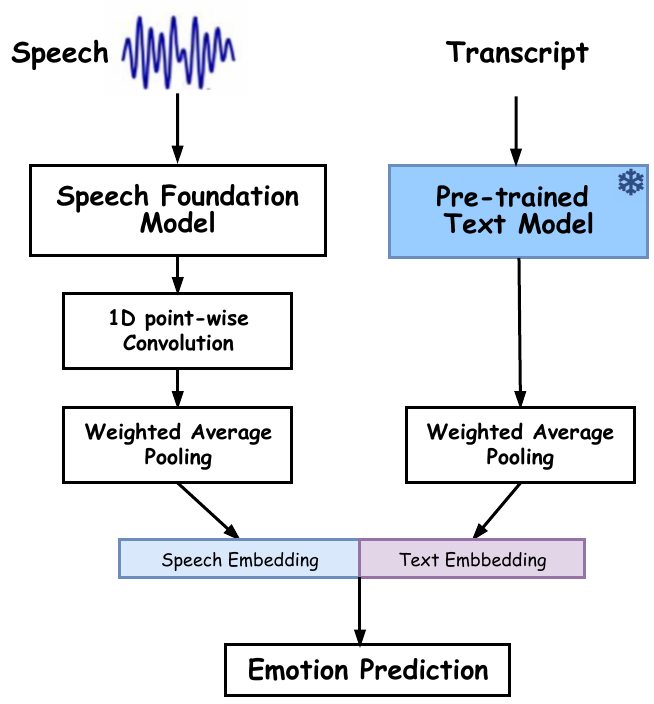}
    \caption{Our proposed multimodal SER framework employs the speech foundation model. Our results show that this simple structure is sufficient to achieve a top-tier performance. MLP stands for multi-layer perceptron.}
    \label{fig:framework}
    \vspace{-4mm}
\end{figure}

In addition to speech modeling, prior studies have identified that textual information benefits SER \cite{li2022fusing, feng2024foundation}. We process the transcript using pre-trained text models to integrate text modeling into the system. Like speech modeling, we apply a weighted average to all encoder outputs, then the temporal averaging. The averaged text output is then concatenated with the averaged speech output to create the multimodal output, which is subsequently passed through a two-layer MLP for classification. Our multimodal architecture is presented in Fig.~\ref{fig:framework}.

\subsection{Learning Objective}

\begin{figure}
    \centering
    \includegraphics[width=0.8\linewidth]{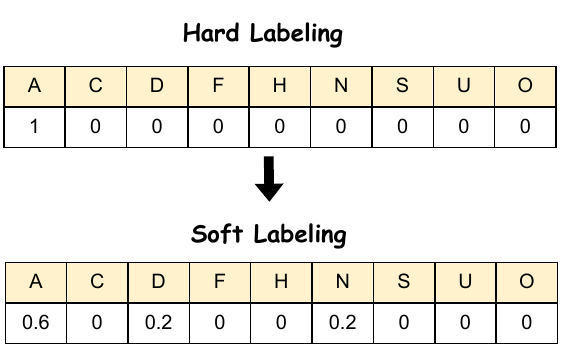}
    \caption{Our proposed soft-labeling approach. This leads to the use of the distribution learning loss (e.g., KL-Divergence).}
    \label{fig:soft_labeling}
    \vspace{-3mm}
\end{figure}

Most prior studies approach SER as a hard-label classification problem. However, speech often conveys multiple emotions simultaneously \cite{Mower2009Interpretingambiguousemotionalexpressions}, and majority-vote hardlabeling fails to capture these multifaceted characteristics \cite{Mower2011AFrameworkforAutomatic}. Moreover, as described in prior research \cite{chou2024minority}, hard labeling can lead to a substantial data loss in training due to the ambiguity of emotions in some speech samples. For example, in the IS25-SER Challenge, 15,932 training speech samples lacked consensus, accounting for around 19\% of the training data. To better represent the complexity of emotions expressed from the speech and use every speech sample in the dataset, \texttt{SAILER} frames SER as a soft labeling problem, as shown in Fig.~\ref{fig:soft_labeling}. Therefore, instead of relying on one-hot classification, we adopt distribution modeling and use KL divergence loss as the learning objective.

\subsection{Data Augmentation}

The modeling choice and learning objective are relatively straightforward to identify in the literature, but determining an effective data augmentation strategy for SER remains challenging. One primary issue in the IS25-SER challenge is the imbalanced training data distribution. Implementing an effective data augmentation method could potentially increase the SER performance by a large margin. \textit{Here, we define majority classes as neutral, happy, sad, and angry, while the remaining emotions are minority classes.} We introduce two novel data augmentations to effectively address the data imbalance issue: annotation dropout and audio mixing.

\begin{figure*}[h]
    \centering
    \includegraphics[width=0.88\linewidth]{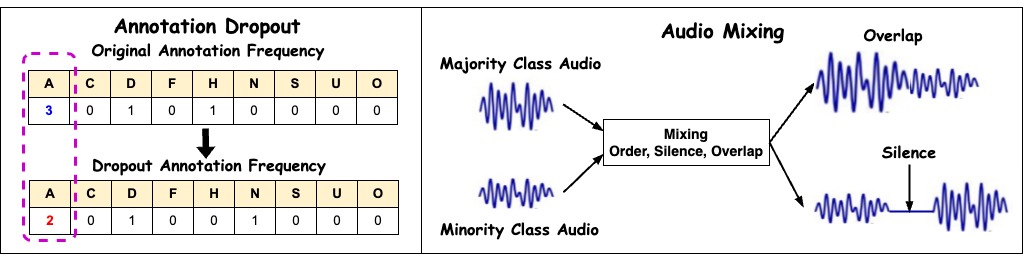}
    \caption{Our proposed data augmentation technique to address data imbalance in SER modeling.}
    \label{fig:augmentation}
    \vspace{-4mm}
\end{figure*}

\vspace{1.5mm}
\noindent \textbf{Annotation Dropout.} Our first data augmentation technique is annotation dropout. \textbf{The key observation we have in (speech) emotion labeling is that the oracle emotion distribution of a speech sample is difficult to obtain in practice.} While each sample in the challenge dataset has at least five annotations, this is still insufficient for estimating its precise emotion distribution $d$. As a result, simple aggregation of annotations introduces inherent noise and biases. To mitigate the noise or biases in this scenario, we randomly drop 20\% of the annotations for each speech sample during training to increase the robustness of the model to a slight distribution shift from $d$ to $\hat{d}$. Given the highly skewed label distribution of the training data, we specifically drop annotations from the majority classes. However, we want to highlight that a more systematic approach would be to drop annotations based on the empirical emotion distributions.

\vspace{1.5mm}
\noindent \textbf{Audio Mixing}. Our second data augmentation technique involves mixing different audio samples in the training dataset to mitigate data imbalance. Specifically, we aim to mix speech samples from majority classes with those from minority classes. For each majority-class speech sample $x_{maj}$, we apply this augmentation with a probability of $p_{a}$. We then select a minority-class speech sample, $x_{min}$, based on an inverse empirical distribution, meaning higher sampling probabilities to less frequently occurring emotion classes. To introduce variability, we randomly determine the order of $x_{maj}$ and $x_{min}$ when mixing. Additionally, we introduce further variability by deciding, via a coin flip, whether to add silence between the samples or create overlapping segments. Finally, we sample a time value $t\in[0, 2]$ to determine the duration of the silence or overlap. Given the emotion distribution of $d_{maj}$ and $d_{min}$, the emotion distribution for the mixed audio $d_{mix} = (d_{maj} + d_{min}) / 2$.

\subsection{Engineering Design Choices to Tackle Class Imbalance}

In addition to the aforementioned design choices, we describe several simple engineering choices that we find helpful in tackling class imbalance in SER training. 

\vspace{1.5mm}
\noindent \textbf{Distribution Re-weighting.} The first choice is to re-weight the emotion distribution. We first estimate the training data's empirical emotion distribution $q$ by aggregating emotion distributions from each training sample. Then, the weight $w^{i}$ for the $i$-th emotion is simply $\frac{1}{q^{i}}$. We subsequently normalize $w$ to have a sum of 1. During training, for each augmented speech sample with distribution $\hat{d}$, the re-weighted distribution for each speech sample becomes $\hat{d}' = \hat{d}\circ w$. We want to underscore that sample re-weighting is unnecessary for validation or testing.

\vspace{1.5mm}
\noindent \textbf{Validation Metrics.} We frequently find that, although a system achieves a decent overall performance as measured by macro-F1 scores, this may be largely driven by the majority classes. Relying solely on the overall macro-F1 scores could lead to worse performance on the test set where the emotion distribution is balanced. Therefore, it is critical to incorporate additional metrics that specifically assess the performance of minority classes. To address this, we introduce the average precision of minority class prediction as an additional validation metric.

\vspace{1.5mm}
\noindent \textbf{Predicting Additional Ground Truth.} Given the additional ground-truth annotations for each speech sample, such as secondary emotions, arousal, valence, and dominance labels, it is worthwhile to explore a simple multi-task learning approach that simultaneously learns the primary emotion distribution along with other affective elements of speech.

\begin{table}
    \centering
    \caption{Dataset statistics of IS25-SER challenge.}
    \vspace{-1.5mm}
    \begin{tabular}{lccc}
        \toprule
         & \textbf{Training} & \textbf{Development} & \textbf{Test} \\
        \midrule
        \textbf{Neutral} & 29,243 & 7,423 & 400 \\
        \textbf{Happy} & 16,717 & 6,344 & 400 \\
        \textbf{Sad} & 6,306 & 2,341 & 400 \\
        \textbf{Disgust} & 1,432 & 542 & 400 \\
        \textbf{Angry} & 6,731 & 5,836 & 400 \\
        \textbf{Contempt} & 2,495 & 1,459 & 400 \\
        \textbf{Fear} & 1,120 & 326 & 400 \\
        \textbf{Surprise} & 1,120 & 987 & 400 \\
        \textbf{Other} & 2,948 & 642 & 0 \\
        \textbf{No Agreement} & 15,932 & 6,061 & 0 \\
        \bottomrule
    \end{tabular}
    \vspace{2mm}
    \label{tab:dataset}
    \vspace{-3.5mm}
\end{table}

\begin{table}
    \centering
    \caption{Unimodal and Multimodal comparisons between two speech foundations models: WavLM large and Whisper-Large V3. The performance is in macro-F1 and Accuracy.}
    \vspace{-2mm}
    \begin{tabular}{ccccc}
        \toprule
         & \multicolumn{2}{c}{\textbf{Unimodal}} & \multicolumn{2}{c}{\textbf{Multimodal}} \\
        & Macro-F1 & Acc & Macro-F1 & Acc \\
        \cmidrule(lr){1-1} \cmidrule(lr){2-3} \cmidrule(lr){4-5}
        \textbf{WavLM Large} & 0.376 & 50.72 & 0.389 & \textbf{55.97} \\
        \textbf{Whisper-Large} & \textbf{0.383} & \textbf{53.67} & \textbf{0.402} & 55.42 \\
        \bottomrule
    \end{tabular}
    \vspace{2mm}
    \label{tab:baseline}
    \vspace{-6mm}
\end{table}

\section{Dataset and Experiment}

\subsection{Dataset}

The IS2025 Emotion Recognition Challenge used the MSP-Podcast dataset v1.12 \cite{lotfian2017building, Naini_2025}. The dataset consists of podcast data from the Internet, including spontaneous speech with natural human emotion expressions. The dataset is annotated with different emotion attributes. The dataset consists of five subsets: the training, development, and three unique test sets. The IS2025 SER challenge uses the test-3 set as the test set where the ground-truth labels have not been made public. Our experiments used the entire training set, which includes nine primary emotion classes: neutral, happy, angry, disgust, sad, surprise, contempt, fear, and others. We also included samples with no agreement on emotion labeling. 

Given that we confirmed with the challenge organizer that no specified rule limited the use of the dataset development set for training, we decided to also include the 'other' and 'no-agreement' samples from the development set as training data in some later challenge submissions. While it may be trivial to improve the current system (by at least 1-2\%) by including more of the remaining development samples, we choose not to do so in order to ensure a consistent finding. Detailed information on the training set and development set is reported in Table~\ref{tab:dataset}.

\vspace{-1mm}
\subsection{Experimental Details}

While we acknowledge that hyperparameter tuning could improve our current system, our primary goal is to show the effectiveness of \texttt{SAILER} with minimal hyperparameter search. Therefore, all of our systems are trained with a learning rate of 0.0005 for 15 epochs. The filter size in the downstream convolutional module is consistently set to 256 across all experiments. We set the maximum of the speech input to 15 seconds and used a fixed seed for all experiments. We base these parameters on our previous work in \cite{feng2023peft} without significant modifications.

We evaluate two speech foundation models: WavLM Large and Whisper Large-V3. Following the challenge baseline, we fine-tune the pre-trained WavLM Large along with the downstream models, whereas for Whisper Large-V3, we fine-tune only the downstream models. For WavLM Large, we apply a weighted average over all encoder outputs, while for Whisper Large-V3, we use only the representations from the last layer. We use the RoBERTa-Large \cite{liu2019roberta} as the pre-trained text model. The limited opportunities for test set submissions make it difficult to systematically report test performance; consequently, we primarily report validation results in most of our experiments. We find that training SER based on Whisper-Large V3 is highly efficient, requiring just 12GB of GPU memory and approximately 15 GPU hours.

\vspace{-1mm}
\section{Results}

\subsection{Do Speech Foundation Models Impact SER?}

As suggested by \cite{feng2024foundation}, we first investigate whether the choice of speech foundation models impacts the SER performance. Specifically, we compare the performance of WavLM Large and Whisper-Large V3 under both speech-only and multimodal conditions that are trained using only the training set. Our results in Table~\ref{tab:baseline} indicate that the selection of the pre-trained model has a notable impact on SER performance, with Whisper-Large V3 consistently outperforming WavLM Large. One plausible reason for this advantage is the larger dataset used to train Whisper-Large V3, which likely yields a more generalizable speech representation for downstream tasks. We wish to highlight that even a simple unimodal WavLM Large solution could yield a test macro-F1 score above 0.37, achieving the top 30 ranks (out of 166 submissions) in the leaderboard.

\begin{table}
    \centering
    \caption{Comparisons of models trained without augmentation to those with audio mixing, annotation dropout, and both combined. Min. mAP indicates the average precision over the minority emotions. All the experiments are based on the multimodal solution that used Whisper-Large V3.}
    \vspace{-2mm}
    \begin{tabular}{lcccc}
        \toprule
        \textbf{Augmentation} & \textbf{Macro-F1} & \textbf{Acc} & \textbf{Min. mAP} \\
        \cmidrule(lr){1-1} \cmidrule(lr){2-4}
        Not Applied & 0.401 & 53.82 & 0.301  \\
        Annotation Dropout & 0.403 & 53.87 & 0.321 \\
        Audio Mixing & 0.404 & 53.42 & \textbf{0.336} \\
        Dropout and Mixing & \textbf{0.406} & \textbf{53.46} & 0.328 \\
        \bottomrule
    \end{tabular}
    \label{tab:augmentation}
    \vspace{-2mm}
\end{table}

\vspace{-2mm}
\subsection{Does Data Augmentation Impact SER?}

Next, we explore the impact of data augmentation on SER performance, comparing results with and without augmentation. Specifically, we evaluate audio mixing, annotation dropout, and their combination within the Whisper-Large multimodal setup, as presented in Table~\ref{tab:augmentation}. We want to highlight that, in this experiment, we included speech samples with 'other' and 'no agreement' labels from the development set as training data. While the results show no improvement in overall accuracy, both augmentation techniques improve SER performance, as measured by the macro-F1 score. Additionally, the minority class average precision indicates that both augmentation techniques are beneficial in improving predictions for minority classes.

\begin{table}
    \centering
    \caption{Comparisons of models trained solely on primary emotions to those including secondary emotions, attribution labels, or a combination of both. Min. mAP indicates the average precision for the minority emotions. All experiments are conducted using a multimodal approach based on Whisper-Large.}
    \footnotesize
    \vspace{-2mm}
    \begin{tabular}{lcccc}
        \toprule
        \textbf{Additional Labels} & \textbf{Macro-F1} & \textbf{Acc} & \textbf{Min. mAP} \\
        \cmidrule(lr){1-1} \cmidrule(lr){2-4}
        Not Included & 0.406 & 53.46 & \textbf{0.328}  \\
        2nd Emotions & 0.409 & 54.26 & 0.322 \\
        Emotion Attributes & 0.407 & 53.42 & 0.305 \\
        2nd Emotions + Attributes & \textbf{0.411} & \textbf{54.53} & 0.316 \\
        \bottomrule
    \end{tabular}
    \label{tab:additional_labels}
    \vspace{-3mm}
\end{table}

\vspace{-2mm}
\subsection{Does Predicting Additional Labels Impact SER?}

This section investigates whether multitask learning with additional emotion annotations, such as secondary emotions or attribute scores (e.g., arousal), benefits SER performance. We conduct our experiments using the same settings as in the data augmentation experiments, where audio mixing and annotation dropout are applied. The results indicate that incorporating additional prediction targets can improve the overall macro-F1 score. However, we observe a decline in performance for minority classes. It is worth noting that a single system that predicts both primary and secondary emotions, as shown in Table~\ref{tab:additional_labels}, achieves a top-15 ranking on the leaderboard. Although we have not yet tested all models listed in Table~\ref{tab:additional_labels}, we anticipate that the best-performing single system might be sufficient to achieve approximately 0.41 macro-F1 on the test set.

\begin{table}
    \centering
    \caption{Comparisons between the best single system and the 3-system ensembles. Our 3-system ensembles achieve a macro-F1 score above 0.41 on the test set.}
    \label{tab:additional_labels}
    \vspace{-2mm}
    \begin{tabular}{lcccc}
        \toprule
        \textbf{System Ensembles} & \textbf{Macro-F1} & \textbf{Acc} & \textbf{Min. mAP} \\
        \cmidrule(lr){1-1} \cmidrule(lr){2-4}
        Best Single System & 0.411 & 54.53 & 0.316  \\
        2-System Ensemble & 0.424 & 56.17 & 0.321 \\
        3-System Ensemble & \textbf{0.431} & \textbf{57.00} & \textbf{0.325} \\
        \bottomrule
    \end{tabular}
    \vspace{-3.5mm}
\end{table}

\vspace{-2mm}
\subsection{How many system ensembles are needed?}

Following our design philosophy of prioritizing simplicity and reproducibility while avoiding complex ensembles, we limit our implementation to a maximum of three systems. Encouragingly, we found that an ensemble of three systems is sufficient to achieve a ranked performance on the leaderboard. Specifically, our ensemble includes system-1, the unimodal WavLM-Large model; system-2, a multimodal Whisper-Large model with multitask learning of secondary emotions; system-3, a multimodal Whisper-Large model with multitask learning of all emotion attributes and secondary emotions. Moreover, we present a two-system ensemble comprising system-1 and system-2 as a reference. While integrating additional systems may further improve SER performance, we leave this exploration for future work.

\section{On Further Improvements}

While our current system shows competitive performance, we highlight several promising and easy-to-prototype directions that researchers can explore for developing next-generation, state-of-the-art SER systems. One is to study pre-trained speech models with emotional speech data like Emotion2Vec \cite{ma2024emotion2vec}.

\vspace{1.5mm}
\noindent \textbf{Learning Objective.} While we use the KL-Divergence as the loss function, a straightforward approach to model the emotion distribution, there are other promising alternatives. For example, prior work by \cite{chou2024minority} demonstrated that soft cross-entropy consistently outperforms KL-Divergence loss.

\vspace{1.5mm}
\noindent \textbf{Audio Mixing.} Prior work in \cite{dang2023emix} has explored similar ideas of mixing audio for SER. However, unlike our method, they mix speech samples with the same emotion to increase the confidence levels of the predictions. Thus, a promising research direction is systematically exploring audio mixing.

\vspace{1.5mm}
\noindent \textbf{Predicting Additional Ground Truth}. Although \texttt{SAILER} considered modeling additional ground-truth labels, such as secondary emotions from the dataset, other speaker-specific information, like gender and age information, has not been explored in this current system. However, emotional expression can differ significantly across speaker groups. A straightforward way to improve our system could include sex and/or age prediction as an additional learning objective.

\section{Conclusion}

In this work, we describe the \texttt{SAILER} framework, a simple and reproducible-friendly SER model for categorized motion prediction in task 1 of the IS25-SER challenge. \texttt{SAILER} considers designs from speech foundation models and learning objectives to data augmentation and engineering choices to tackle imbalanced data. Experimental results show that \texttt{SAILER} is highly competitive in the IS25-SER challenge, achieving top-tier performance with minimum system complexity.

\section{Acknowledgment}
We gratefully acknowledge support from IARPA ARTS (award number 140D0424C0067, JHU subcontract) from the Office of the Director of National Intelligence and NSF Grant (SCH with award number 2204942).

\bibliographystyle{IEEEtran}
\bibliography{mybib}

\end{document}